# Tuning Leaky Nanocavity Resonances - Perturbation Treatment

*Michael Shlafman[*], Igal Bayn and Joseph Salzman*

Department of Electrical Engineering and Microelectronics Research Center,

Technion, Israel Institute of Technology, Haifa 32000, Israel

* Corresponding author. E-mail: mikester@tx.technion.ac.il

*Adiabatic frequency tuning of finite-lifetime-nanocavity electromagnetic modes affects also their quality-factor (Q). Perturbative Q change resulting from (real) frequency tuning, is a controllable parameter. Here, the influence of dielectric constant modulation (DCM) on cavity resonances is presented, by first order perturbation analysis for a 3D cavity with radiation losses. Semi-analytical expressions for DCM induced cavity mode frequency and Q changes are derived. The obtained results are in good agreement with numerical calculations.*

Recent progress in photonic crystal ultra-high-$Q$ cavity design and fabrication[1-3] makes them very attractive for implementation in the fields of integrated optics (routing and filtering), basic light-matter interaction (cavity quantum electrodynamics) or in quantum-information technologies (single-photon-detectors and quantum-bit-architecture). *Dynamic photonic tuning* is the situation in which an electromagnetic cavity is modulated while the photonic mode is inside the cavity. Dynamic tuning of an ultra-high-$Q$ nanocavity would be of a significant advantage, since this allows active tuning of both cavity mode decay and frequency, thus providing a dynamic control mechanism over cavity-waveguide, cavity-cavity and light matter interactions in the nano-scale.[4-7] The realization of dynamic tuning may rely on several physical mechanisms, such as temperature-tuning, non-linear effects, or free carrier induced modulation of the dielectric constant. In the latter a significant progress has been reported recently.[8] In a finite quality-factor ($Q$) nanocavity (with radiation losses), the modal fields are resonances, with complex eigenfrequencies. Perturbative tuning affects *both the real and the imaginary part of the frequency.* Thus, it seems timely to seek for a basic understanding of the parameter space of (complex) frequency perturbative tuning. As a result of the tuning, the system is non-stationary and the frequency is not conserved (a modal field can "change its color" before leaving the cavity). One figure of merit in a quantum electrodynamics context can be the number of detuned line widths, or frequency tunability within the mode bandwidth. A second figure of merit is the degree of $Q$-switching capability within a (real) frequency range. For that, estimation of both the real frequency *and* the $Q$ modulation are needed. The objective of this work is to provide a simple semi-analytical description of the nanocavity mode *complex frequency* modulation via dielectric constant tuning.

The effects of perturbative changes in the dielectric constant of a *lossless* electromagnetic resonator and their influence on an eigenmode frequency are known for some time.[9] Perturbation theory for lossless (closed) systems with only real eigenfrequencies was extensively described in quantum mechanics[10-12] and in electromagnetic theory[13,14]. Obviously the operator describing the wave equation for such systems is Hermitic and the eigenmodes are orthonormal and complete. These are key properties for the application of perturbation theory. In contrast to these systems, optical nanocavities in a dielectric or semiconductor structure with finite dimensions, exhibit a finite $Q$ due to *radiative losses*, characteristic to open systems. The Hamiltonian operator describing such systems is non-Hermitic, the modes are not stationary (leaky), hence, not orthonormal. For that reason, traditional perturbation formulation is not applicable in those cases. The perturbation treatment in *open* systems was introduced by using a two-component eigenfunction.[15,16] Bi-orthonormality for non-Hermitian systems was also considered for the application of perturbation theory.[17,18] However, in previous studies addressing leaky modes, only scalar fields in one dimension were analysed. In nanocavity tuning (such as in photonic crystal cavities with a local change in dielectric constant) it is essential to account for radiation losses (with "free space" propagation) of 2D or 3D structures, and for the vector field character of the electromagnetic modes. Here, normalizable leaky modes (NLM's) in a finite volume are introduced to treat 3D electromagnetic problems with vector fields[19]. By applying perturbation theory on the NLM's, the effects on a resonant mode frequency and $Q$ due to slight changes in the dielectric constant of the system are predicted. For a lossless system, adiabatic tuning is properly characterized by a time independent perturbation formalism.[12] We consider the "adiabatic" perturbation, and assume that, similarly to conservative systems, time independent perturbation is appropriate to describe the eigenfrequency evolution. As an example, these results are compared to numerical calculations of a photonic crystal nanocavity (in 2D).

Let $\mathbf{E}$ and $\mathbf{H}$ be the complex electric and magnetic fields related by the Maxwell equations. The field $\mathbf{M}$ is defined by $\mathbf{M} = \mu_0 \nabla \times \mathbf{H}$ where $\mu_0$ is the vacuum permeability. Next the two-component vector and the Hamiltonian matrix operator are defined by

$$(a) \qquad |\phi\rangle = \begin{pmatrix} \mathbf{E} \\ \mathbf{M} \end{pmatrix} \qquad\qquad (b) \qquad \hat{\mathbf{\Theta}} \equiv i \begin{pmatrix} 0 & \dfrac{c^2}{\varepsilon_r(\mathbf{r})} \\ -\nabla \times \nabla \times & 0 \end{pmatrix} \qquad (1)$$

where $c$ is the speed of light, $\varepsilon_r$ is the dielectric constant and $\mathbf{r}$ is the coordinate in space. Under such definitions the electromagnetic wave equation may be cast in a time evolution Hamiltonian form: $\hat{\mathbf{\Theta}}|\phi\rangle = i\frac{\partial}{\partial t}|\phi\rangle$.[20] The nanocavity is considered to be a part of a finite size dielectric system of volume $V$ embedded in free space (or a different homogeneous dielectric constant environment) where the boundary surface of $V$ ($S$) experiences a discontinuity in the dielectric constant. The two component vector space $|\phi\rangle$ is the mathematical entity of the Hamiltonian problem. Physically, the leaky cavity modes can be found by solving the wave equation under the outgoing wave boundary condition for the field $\mathbf{E}$

$$\oiint_S \left[ \frac{n^{out}}{c} \frac{\partial}{\partial t} \mathbf{E} + (\nabla \times \mathbf{E}) \times \hat{\mathbf{n}} \right] ds = 0 \qquad (2)$$

where $n^{out} = \sqrt{\varepsilon_r(\mathbf{r})} = const : \mathbf{r} \notin V$ [21]. The solutions form an infinite discrete set of eigenvectors (eigenfunctions) $|\phi_n\rangle$ with corresponded complex eigenvalues (eigenfrequencies) $\omega_n$ for $n = 0, \pm 1, \pm 2, \ldots$. For the harmonic time evolution ($\exp(-i\omega t)$) we obtain that all $\omega_n$ have a negative imaginary part, indicating oscillatory and time decaying modes. This set of modes comprises the leaky modes which under the usual inner product are non orthogonal and non normalizable. To overcome this, we define the following bilinear form[22]

$$\langle \phi_n | \phi_k \rangle \equiv i \iiint\limits_V \left( \mathbf{E}^n \cdot \mathbf{M}^k + \mathbf{E}^k \cdot \mathbf{M}^n \right) d^3\mathbf{r} + i \frac{n^{out}}{c} \oiint\limits_S \left( \mathbf{E}^n \cdot \mathbf{E}^k \right) ds \qquad (3)$$

where $\mathbf{E}^n$, $\mathbf{M}^n$ and $\mathbf{E}^k$, $\mathbf{M}^k$ are the two components of vectors $|\phi_n\rangle$ and $|\phi_k\rangle$ respectively. Equation 3 is the equivalent to an inner product for our purpose, under which the operator $\hat{\Theta}$ is symmetric while the leaky modes are both orthonormal and normalizable (NLM's). Thus, we have defined what can be regarded as a "Hermitic domain" of the operator $\hat{\Theta}$ analogous to Hermiticity in conservative systems.

The presented framework makes possible the use of perturbation theory on electromagnetic systems with radiation losses applicable to nanocavities. Assuming the dielectric constant of the system is $\varepsilon_r(\mathbf{r}) = \varepsilon_r^0(\mathbf{r}) + \varepsilon_r'(\mathbf{r})$ where $\varepsilon_r^0(\mathbf{r})$ is the dielectric constant spatial profile of the unperturbed system and $\varepsilon_r'(r)$ is the spatial profile of the dielectric constant perturbation with $\varepsilon_r'(\mathbf{r}) \ll \varepsilon_r^0(\mathbf{r})$, the Hamiltonian $\hat{\Theta}$ may be split into two parts

$$\hat{\Theta} = \left(\hat{\Theta}^0 + \hat{\Theta}'\right) = i\begin{pmatrix} 0 & \frac{c^2}{\varepsilon_r^0(\mathbf{r})} \\ -\nabla \times \nabla \times & 0 \end{pmatrix} + i\begin{pmatrix} 0 & \frac{-c^2 \varepsilon_r'(\mathbf{r})}{\varepsilon_r^0(\mathbf{r})^2} \\ 0 & 0 \end{pmatrix} \qquad (4)$$

where $\hat{\Theta}^0$ is the operator describing the unperturbed system and $\hat{\Theta}'$ describes only the dielectric perturbation which is small compared to $\hat{\Theta}^0$. From this point, following the first order perturbation theory, we obtain[23]

$$\left|\phi_n\right\rangle = \left|\phi_n^{(0)}\right\rangle + \sum_k \frac{\left\langle \phi_k^{(0)} \middle| \hat{\Theta}' \middle| \phi_n^{(0)} \right\rangle}{\left\langle \phi_k^{(0)} \middle| \phi_k^{(0)} \right\rangle \left(\omega_n^{(0)} - \omega_k^{(0)}\right)} \left|\phi_k^{(0)}\right\rangle \qquad (5)$$

$$\Delta\omega_n = \frac{\left\langle \phi_n^{(0)} \middle| \hat{\Theta}' \middle| \phi_n^{(0)} \right\rangle}{\left\langle \phi_n^{(0)} \middle| \phi_n^{(0)} \right\rangle} \qquad (6)$$

where $\left|\phi_n^{(0)}\right\rangle$ and $\omega_n^{(0)}$ are the two-component NLM's and complex eigenfrequencies of the unperturbed system respectively, and $\left|\phi_n\right\rangle$, $\omega_n$ are the instantaneous eigenmodes and eigenfrequencies of the new perturbed system ($\omega_n = \omega_n^{(0)} + \Delta\omega_n$) to the first order approximation. Note that the familiar looking eqs 5 and 6 receive entirely new interpretation under the new bilinear form defined in eq 3, unlike the result of the conventional electromagnetic (CEM) perturbation theory[14,20]. Elaborating eq 6 leads to

$$\Delta\omega_n = -\left(\frac{\omega_n^{(0)}}{c}\right)^2 I \iiint_V \varepsilon_r'(\mathbf{r}) \mathbf{E}^n \cdot \mathbf{E}^n d^3\mathbf{r} \qquad (7)$$

where $I = \left(1/\left\langle \phi_n^{(0)} \middle| \phi_n^{(0)} \right\rangle \right)$ which is equal to one with the units $[ms^{-1}V^2]$ when the mode $\left| \phi_n^{(0)} \right\rangle$ is normalized. From eq 7 it follows that induced material perturbation of various spatial profiles may yield different complex $\Delta\omega_n$ values depending on the overlap between $\varepsilon_r'(\mathbf{r})$ and the spatial profile of the complex value $\boldsymbol{w}^n \equiv \mathbf{E}^n \cdot \mathbf{E}^n$ of the normalized mode. Equation 7 can also be written as

$$\Delta\omega_n = \frac{1}{2}\,\omega_n^0\,\frac{-\iiint\limits_V \varepsilon_r'(\mathbf{r})\mathbf{E}^n \cdot \mathbf{E}^n d^3\mathbf{r}}{\iiint\limits_V \varepsilon_r^0(\mathbf{r})\mathbf{E}^n \cdot \mathbf{E}^n d^3\mathbf{r} + i\dfrac{n^{out}}{2}\dfrac{c}{\omega_n^0}\oiint\limits_S \mathbf{E}^n \cdot \mathbf{E}^n ds}\,. \tag{8}$$

The most notable difference between eq 8 and the expression of $\Delta\omega$ in conservative system formalism[9] is the existence of a surface term in the denominator. Another difference is that in the integrals (such as $\boldsymbol{w} \equiv \iiint_V \left(\varepsilon_r'(\mathbf{r})\mathbf{E}^n \cdot \mathbf{E}^n\right) d^3\mathbf{r}$ in the numerator), both of the electric field terms $\mathbf{E}^n$ appear *without conjugation* (unlike in former treatments[9] where $\left|\mathbf{E}^n\right|^2$ $(= \mathbf{E}^n \cdot \mathbf{E}^{n*})$ is used). The last difference is related to the integration volume which, unlike in conservative systems, is over the volume $V$ enclosing the dielectric system and defined by its boundaries.

Concerning the cavity $Q$, and its perturbative change, using the relation $Q = \left(-\operatorname{Re}\{\omega\}/2\operatorname{Im}\{\omega\}\right)$, we obtain

$$\frac{\Delta Q_n}{Q_n^0} = \left[ -\frac{Q_n^0 \boldsymbol{w}_i - \frac{1}{2}\boldsymbol{w}_r}{\left(\dfrac{c^2}{2\left|\omega_n^{(0)}\right|\hat{\mathbf{i}}}\right) + Q_n^0 \boldsymbol{w}_i - \boldsymbol{w}_r} \right] \tag{9}$$

where $Q_n^0$ and $\Delta Q_n$ are the quality factor of eigenmode $n$ in the unperturbed system and the change in the quality factor induced by the perturbation, respectively, while $\boldsymbol{w}_r = \mathrm{Re}\{\boldsymbol{w}\}$ and $\boldsymbol{w}_i = \mathrm{Im}\{\boldsymbol{w}\}$. Equations 7 and 9 can be applied, to optimize either the cavity design, or the spatial distribution of the perturbation, in order to obtain the desirable $\Delta\omega$ with a certain degree of separate control over its real and imaginary values, (i.e. frequency and $Q$). This derivation allows a semi-analytical prediction of the dielectric constant modulation effect on the system mode without the need of complex and time consuming numerical calculations.

In order to test the accuracy of the perturbative expressions, the values of $\Delta\omega$ and $\Delta Q$ predicted by eqs 7 and 9 where compared to the CEM perturbation theory predicted values and to the results of a finite difference time domain (FDTD)[24] simulation in a two-dimensional (2D) nanocavity example. The simulated structure is depicted in Figure 1a. The nanocavity consists of a finite size rectangular 2D photonic crystal with a $7 \times 7$ periodic lattice of air holes ($\varepsilon_r = 1$) in silicon ($\varepsilon_r = 12.25$). The reduced hole size in the middle of the structure defines the nanocavity. The unit cell size is $a$, the radius of a regular lattice and nanocavity air holes are $r = 0.48a$ and $r' = 0.15a$, respectively. The structure exhibits a bandgap in the frequency range of $0.2425(2\pi c/a) \leq \omega \leq 0.3115(2\pi c/a)$ for the TM polarization ($\mathbf{E} = E_z \hat{\mathbf{z}}$) and supports three cavity modes. The highest cavity NLM $\left|\phi_{III}^{(0)}\right\rangle$ is analyzed. This NLM is characterized by $\omega_{III}^0 = (0.282732 - 0.000168608i)(2\pi c/a)$ and $Q_{III}^0 = 838.429$, (unperturbed simulation values). The real and imaginary parts of the NLM's electric field ($\mathbf{E}^{III}$) and its square ($\boldsymbol{w}^{III}$) normalized according to $\left\langle\phi_{III}^{(0)}\middle|\phi_{III}^{(0)}\right\rangle = 1$ are shown in Figures 1b-1e with normalized intensity[25]. We define two perturbation regions. Region 1 in the middle of the structure near the cavity, marked by dotted green line and region 2 in the cavity surroundings, marked by dashed purple line (see Figure 1a).

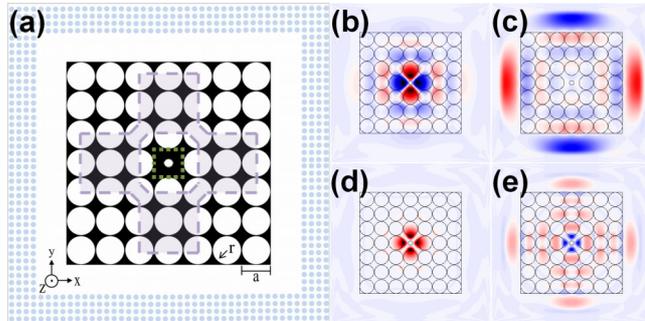

**Figure 1.** (a) Simulated structure. The Perfectly Matched Layer (PML) is marked by light blue dots region at the borders. Green dotted line marks region 1 of the perturbation. Purple dashed line marks region 2 of the perturbation. (b)-(e) Highest frequency normalized cavity mode $\left|\phi_{III}^{(o)}\right\rangle$. (b) $\mathrm{Re}\left\{\mathbf{E}^{III}\right\}$ .(c) $\mathrm{Im}\left\{\mathbf{E}^{III}\right\}$ .(d) $\mathrm{Re}\left\{\boldsymbol{w}^{III}\right\}$ .(e) $\mathrm{Im}\left\{\boldsymbol{w}^{III}\right\}$ [25].

At first a dielectric constant perturbation was introduced to the silicon in region 1. For perturbations of $\varepsilon_r'/\varepsilon_r^0 = 0.0001$, 0.001 and 0.01, (*pert's* 1-3) where $\varepsilon_r'$ and $\varepsilon_r^0$ are the dielectric region perturbation, and unperturbed values, respectively. The "reference" values for $\Delta\omega_{III}$ and $\Delta Q_{III}$ are obtained by the numerical simulations for each one of the three perturbations. The values of $\Delta\omega_{III}$ and $\Delta Q_{III}$ were then estimated using the NLM perturbation theory and the CEM perturbation theory denoted by $\Delta\omega_{III}^{NLM}$ ,$\Delta Q_{III}^{NLM}$ and $\Delta\omega_{III}^{CEM}$ ,$\Delta Q_{III}^{CEM}$, respectively[26]. Table 1 shows the values of $\Delta\omega_{III}$ and $\Delta Q_{III}$ for each one of the three perturbations (simulation obtained and predicted values) and the corresponding relative prediction errors $\delta_Q^{NLM} \equiv$

$\left( \left| \Delta Q_{III} - \Delta Q_{III}^{NLM} \right| / \left| \Delta Q_{III} \right| \right)$ and $\delta_{\omega-re}^{NLM} \equiv \left( \left| \mathrm{Re}\{\Delta\omega_{III}\} - \mathrm{Re}\{\Delta\omega_{III}^{NLM}\} \right| / \left| \mathrm{Re}\{\Delta\omega_{III}\} \right| \right)$ for the

NLM perturbation method and similarly $\delta_{\omega-re}^{CEM}$ for the CEM perturbation method.

**Table 1.** Frequency and $Q$ changes of the highest nanocavity mode obtained by an FDTD simulation and predicted by the Normalizable Leaky Modes and Conventional Electro-Magnetic perturbation treatments accompanied by the relative errors in comparison to the FDTD results.

|  | pert 1 | pert 2 | pert 3 |
|---|---|---|---|
| $\varepsilon_r' / \varepsilon_r^0$ | 0.0001 | 0.001 | 0.01 |
| $\Delta\omega_{III}$ $(2\pi c/a)$ | $-6.75 \times 10^{-6} + 5.035i \times 10^{-8}$ | $-6.7501 \times 10^{-5} + 5.0237i \times 10^{-7}$ | $-6.7639 \times 10^{-4} + 4.9097i \times 10^{-6}$ |
| $\Delta\omega_{III}^{NLM}$ $(2\pi c/a)$ | $-6.8949 \times 10^{-6} + 4.9509i \times 10^{-8}$ | $-6.8948 \times 10^{-5} + 4.9508i \times 10^{-7}$ | $-6.894 \times 10^{-4} + 4.9503i \times 10^{-6}$ |
| $\Delta\omega_{III}^{CEM}$ $(2\pi c/a)$ | $-7.1481 \times 10^{-6} + 4.2628i \times 10^{-9}$ | $-7.1481 \times 10^{-5} + 4.2628i \times 10^{-8}$ | $-7.1477 \times 10^{-4} + 4.2626i \times 10^{-7}$ |
| $\Delta Q_{III}$ | 0.2304 | 2.3048 | 23.080 |
| $\Delta Q_{III}^{NLM}$ | 0.2258 | 2.2640 | 23.255 |
| $\Delta Q_{III}^{CEM}$ $^a$ | ~0 | ~0 | ~0 |
| $\delta_{\omega-re}^{NLM}$ | 2.147% | 2.130% | 1.924% |
| $\delta_{\omega-re}^{CEM}$ | 5.899% | 5.882% | 5.675% |
| $\delta_Q^{NLM}$ $^b$ | 2.003% | 1.767% | 0.754% |

$^a$ The elaborated expression for $\Delta\omega_{III}^{CEM}$ is of the form $\Delta\omega_{III}^{CEM} = \omega_{III}^{(0)} g$ where $g = g\left(\mathbf{E}_{III}, \mathbf{H}_{III}, \varepsilon_r^0, \varepsilon_r'\right)$ is found to be almost purely real in the examined case. Hence, the ratio $\mathrm{Re}\{\omega_{III}\} / \mathrm{Im}\{\omega_{III}\}$ suffers negligible change, and being proportional to the $Q$, all values of $\Delta Q_{III}^{CEM}$ are nearly zero as well.

$^b$ The values of $\delta_Q^{CEM}$ are omitted since the conventional perturbation theory fails to predict the $Q$ correctly.

As it is shown in the table, the NLM method yields a better approximation for $\text{Re}\{\Delta\omega\}$ than the CEM perturbation method. Out of these two methods only the NLM is capable to predict changes in $\Delta Q$ with a reasonable precision (relative error of about 2% or less). The authors are unaware of other techniques for approximating perturbative changes in $\Delta Q$ to compare with.

Furthermore, according to eq 7, applying the perturbation in a region where $\text{Im}\{\boldsymbol{w}^{III}\}$ is *negative* would increase the imaginary part of the eigenfrequency of the mode, i.e. would increase the $Q$. As one can observe in Figure 1e, region 1 is exactly such a region. In the same way because $\text{Re}\{\boldsymbol{w}^{III}\}$ is *positive* almost everywhere in volume $V$, nearly any positive perturbation to the dielectric constant will decrease the real part of the eigenfrequency, as readily observed in Table 1. Hence, one could expect that applying the perturbation in a region in which $\text{Im}\{\boldsymbol{w}^{III}\}$ is *positive* will change the $Q$ in the opposite direction. Such example is presented as well. Applying perturbation of $\varepsilon_r'/\varepsilon_r^0 = 0.01$ in region 2 (*pert* 4) causes the real part of the eigenfrequency to drop $\Delta\omega_{III} = (-1.6028\times10^{-4} - 2.98225i\times10^{-6})$ and the $Q$ to decrease $\Delta Q_{III} = (-15.039)$. These examples show that there is certain degree of freedom in controlling separately the frequency and the $Q$, by a proper choice of the geometrical distribution of the dielectric constant perturbation. In a more general sense, the present framework makes the exploration of the parameter space in tuning of leaky cavity resonances possible. A different choice of the dielectric structure would alter the spatial profile of $\boldsymbol{w}^n$ which, in turn, is responsible for the system's resonance eigenfrequency sensitivity to a dielectric perturbation, in strength ($\varepsilon_r'/\varepsilon_r^0$) and spatial distribution. This may be an important step towards better capabilities of nanocavity design and resonance tuning by a dielectric perturbation.

The parameter space of complex eigenfrequency response to a dielectric perturbation is further investigated, based on eq 9, by defining a "*Q-response*" function $f_Q^{III} \equiv \left( \Delta Q_{III} / Q_{III}^0 \right)$. For a free choice of $w_r$ and $w_i$ values the surface $f_Q^{III} \left( w_r, w_i \right)$ is depicted in Figure 2a (the areas close to singularity were removed). Two planes, at $f_Q^{III} = 0$ (zero $Q$ change) and $f_Q^{III} = (-1)$ (total nullification of $Q$) are depicted in light gray and dark gray, respectively. All four tested perturbations in the nanocavity example concentrate very close to the black circle in the graph where $f_Q^{III}$ deviates only slightly from zero. In the example studied here, the possible values of $w_r$ and $w_i$ are very limited[27] and $f_Q^{III}$ in this range is almost linear and negligibly dependent on $w_r$ (see Figure 2b). The four calculated perturbations are marked by the four points in Figures 2b, 2c. The first three points above the $f_Q^{III} = 0$ level and the last one below it. Figure 2c presents $f_Q^{III}$ as a function of $w_i$ only for the selected values of $w_r$ corresponding to the actual tested perturbations.

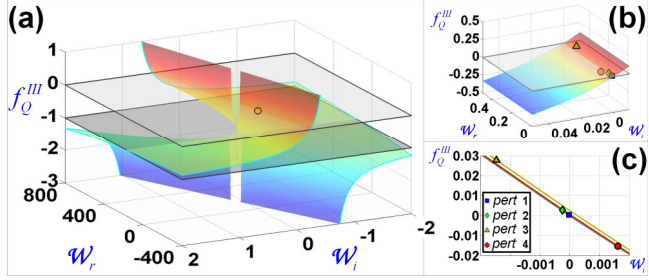

**Figure 2.** (a) The ratio $f_Q^{III}(\mathcal{W}_r, \mathcal{W}_i) \equiv (\Delta Q_{III}/Q_{III}^0)$ in the wide range. (b) $f_Q^{III}(\mathcal{W}_r, \mathcal{W}_i)$ in the relevant range with the specific perturbation marked on it. (c) $f_Q^{III}(\mathcal{W}_i)$ with the four tested perturbations.

One important problem with high-$Q$ nanocavities is the difficulty of out-coupling of excitations from the cavity. One would like a $Q$-switch, to change the cavity Q from high to low in order to release the electromagnetic cavity mode to the external world, or to do the opposite, to increase the resonator's $Q$ to such an extent that the confined light mode which is about to leave the cavity would be trapped inside it for additional period of time.[7] As seen from Figure 2, for the first task one need a high positive value of $\mathcal{W}_i$, and for the second task, a negative value of $\mathcal{W}_i$ which is close to the surface singularity. In the presented examples only slight changes in $Q$ are possible, but, in other systems the situation can be different. Also note that $Q_n^0$ is a multiplicand of $\mathcal{W}_i$ in eq 9, thus, perturbations in ultra-high-$Q$ systems may produce substantial relative changes in the $Q$. However, the spatial distribution of $\text{Im}\{\boldsymbol{w}^n\}$ may neutralize the effect of a very high $Q$ in such systems. These considerations, related to the parameter space for a perturbative change in the eigenfrequency of nanocavities with radiation losses may serve as a design framework for the optimization in both:

frequency tuning and $Q$-switching. Higher order perturbation theory, and quantum gate structures can be studied based on the formalism presented here.[28]

In summary, we have presented the Normalizable Leaky Mode formalism for electromagnetic vector fields in 3D. Then, based on this formalism perturbation theory to electromagnetic waves confined in a dielectric nanocavity with radiation losses was applied, formulas for the changes in the frequency and $Q$ induced by small perturbations of the dielectric constant spatial distribution were derived. A $Q$-response function was defined for investigation of the effect of spatial distribution of the dielectric perturbation on the $Q$ of the nanocavity. Under some conditions slight changes in the dielectric function may strongly affect the $Q$ while slightly affecting the frequency of an eigenmode and vice-versa.[28]

symmetric by definition, bilinear map doesn't have to be, and in the presented case it is *just* symmetric.

23. We define $\left\langle \psi \middle| \hat{\boldsymbol{\Theta}}' \middle| \varphi \right\rangle \equiv \left\langle \psi \middle| \hat{\boldsymbol{\Theta}}' \varphi \right\rangle$, where $\hat{\boldsymbol{\Theta}}' \varphi$ is the matrix by vector multiplication (see eqs 1 and 4).

24. Meep. http://ab-initio.mit.edu/wiki/index.php/Meep (13.10.2009), free finite-difference time-domain (FDTD) simulation software package developed at MIT to model electromagnetic systems.

25. The absolute peak value in the computational region of $\mathrm{Re}\left\{\mathbf{E}^{III}\right\}$ is roughly 23 times larger than that of $\mathrm{Im}\left\{\mathbf{E}^{III}\right\}$. Similarly the absolute peak value in the computational region of $\mathrm{Re}\left\{\boldsymbol{w}^{III}\right\}$ is roughly 172 times larger than that of $\mathrm{Im}\left\{\boldsymbol{w}^{III}\right\}$.

26. Because we deal with leaky modes which extend to the whole universe and even diverge in amplitude at infinity, the $\Delta\omega_{III}^{CEM}$ and $\Delta Q_{III}^{CEM}$ are actually not defined as is, since the 'bra' and 'ket' operations involve integration over the whole space, in contrary to the NLM formalism which is restricted to volume $V$ by definition. Yet for the purpose of applying the CEM perturbation theory and since the cavity mode is confined, we treat it as a conservative mode and restrict the integration only to volume $V$ at the borders of which the mode's field has undergone substantial attenuation compared to its absolute maximum at the vicinity of the cavity.

27. In the examples studied here, even under a very large material modulation of $\varepsilon_r' = 0.15$ ($\varepsilon_r' / \varepsilon_r^0 \cong 0.01$) at every desired point in the volume $V$ (including air holes), the possible values of $\mathcal{W}_r$ and $\mathcal{W}_i$ are limited to $-0.422 \cdot 10^{-3} \leq \mathcal{W}_r \leq 2.341$ and $9.615 \cdot 10^{-3} \leq \mathcal{W}_i \leq 0.0322$ respectively.

28. If, upon cavity design and mode calculation, $\mathcal{W}_r$ and $\mathcal{W}_i$ receive the adequate magnitude and sign, desired response in $\Delta\omega$ or $\Delta Q$ can be obtained. This will be shown elswere (Shlafman, M.; Salzman, J. unpublished).

## Figure Captions

**Figure 1.** (a) Simulated structure. The Perfectly Matched Layer (PML) is marked by light blue dots region at the borders. Green dotted line marks region 1 of the perturbation. Purple dashed line marks region 2 of the perturbation. (b)-(e) Highest frequency normalized cavity mode $\left|\phi_{III}^{(o)}\right\rangle$. (b) $\mathrm{Re}\left\{\mathbf{E}^{III}\right\}$ .(c) $\mathrm{Im}\left\{\mathbf{E}^{III}\right\}$ .(d) $\mathrm{Re}\left\{\boldsymbol{w}^{III}\right\}$ .(e) $\mathrm{Im}\left\{\boldsymbol{w}^{III}\right\}$ [25].

**Figure 2.** (a) The ratio $f_Q^{III}\left(\boldsymbol{\mathcal{W}}_r,\boldsymbol{\mathcal{W}}_i\right)\equiv\left(\Delta Q_{III}/Q_{III}^0\right)$ in the wide range. (b) $f_Q^{III}\left(\boldsymbol{\mathcal{W}}_r,\boldsymbol{\mathcal{W}}_i\right)$ in the relevant range with the specific perturbation marked on it. (c) $f_Q^{III}\left(\boldsymbol{\mathcal{W}}\right)$ with the four tested perturbations.

## Table Caption and Footnotes

**Table 1.** Frequency and $Q$ changes of the highest nanocavity mode obtained by an FDTD simulation and predicted by the Normalizable Leaky Modes and Conventional Electro-Magnetic perturbation treatments accompanied by the relative errors in comparison to the FDTD results.

[a] The elaborated expression for $\Delta\omega_{III}^{CEM}$ is of the form $\Delta\omega_{III}^{CEM}=\omega_{III}^{(0)}g$ where $g=g\left(\mathbf{E}_{III},\mathbf{H}_{III},\varepsilon_r^0,\varepsilon_r'\right)$ is found to be almost purely real in the examined case. Hence, the ratio $\mathrm{Re}\left\{\omega_{III}\right\}/\mathrm{Im}\left\{\omega_{III}\right\}$ suffers negligible change, and being proportional to the $Q$ all values of $\Delta Q_{III}^{CEM}$ are nearly zero as well.

[b] The values of $\delta_Q^{CEM}$ are omitted since the conventional perturbation theory fails to predict the $Q$ correctly.